\documentclass[review]{elsarticle}

\usepackage{lineno,hyperref}

\usepackage{amsmath, amsthm, amsfonts}
\usepackage{graphicx}
\usepackage{mathtools}
\usepackage{url}
\usepackage{array}
\usepackage{multirow}
\usepackage{caption}
\usepackage{subcaption}
\usepackage{makecell}
\usepackage{color,soul}
\usepackage{float}

\usepackage[flushleft]{threeparttable} % http://ctan.org/pkg/threeparttable
\usepackage{booktabs,caption}

\makeatletter
\newcommand\footnoteref[1]{\protected@xdef\@thefnmark{\ref{#1}}\@footnotemark}
\makeatother

\theoremstyle{definition}
\newtheorem{defn}{Definition}[section]
\journal{Journal of Information Sciences}

\begin{document}

\begin{frontmatter}
\title{sCAKE: Semantic Connectivity Aware Keyword Extraction}

\author[mainaddress]{Swagata Duari\corref{mycorrespondingauthor}}
\cortext[mycorrespondingauthor]{Corresponding author}
\ead{sduari@cs.du.ac.in}

\author[mainaddress]{Vasudha Bhatnagar}
\address[mainaddress]{Department of Computer Science, University of Delhi, India}

\begin{abstract}
Keyword Extraction is an important task in several text analysis endeavours.  %In last two decades, graph-based methods for automatic keyword extraction have seen impressive developments. 
In this paper, we present a critical discussion of the issues and challenges in graph-based keyword extraction methods, along with comprehensive empirical analysis. We propose a parameterless method for constructing graph of text that captures the contextual relation between words. A novel word scoring method is also proposed based on the connection between concepts. We demonstrate that both proposals are individually superior to those followed by the sate-of-the-art graph-based keyword extraction algorithms. Combination of the proposed graph construction and scoring methods leads to a novel, parameterless keyword extraction method (sCAKE) based on semantic connectivity of words in the document.

Motivated by limited availability of NLP tools for several languages, we also design and present a language-agnostic keyword extraction (LAKE) method. We eliminate the need of NLP tools by using a statistical filter to identify candidate keywords before constructing the graph. We show that the resulting method is a competent solution for extracting keywords from documents of languages lacking sophisticated NLP support.
\end{abstract}


\end{frontmatter}

%\linenumbers

%%%%%%%%%%%%%%%% INTRODUCTION %%%%%%%%%%%%%%%%%%%%
%\input{mycontent/Introduction}
\section{Introduction}
\label{sec:intro}

Modern search engines and document databases are tasked with identifying and locating information with high efficiency. This is typically done using {\it keywords} - a small set of relevant and important terms that sufficiently describe the given document. \textit{Keyword extraction} task is associated with extracting such terms from a document. According to Ohsawa et al.~\cite{ohsawa1998keygraph}, assigning representative terms to a document is a process called indexing and the terms assigned are known as \textit{keywords}. Indexing significantly reduces the human effort in sifting through vast amounts of information. With monotonically growing repositories of digital documents, study of automatic keyword extraction methods has attracted serious attention \cite{boudin2018unsupervised,carpena2009level,carretero2013improving,florescu2017position,hulth2003improved,litvak2011degext,matsuo2001keyworld,mihalcea2004textrank,ortuno2002keyword,rousseau2015main,yu2017ci}. Effective keyword extraction methods lead to improved indexing in massive text repositories, thereby enhancing the quality of retrieved search results.

Automatic keyphrase extraction is a natural extension of keyword extraction problem, where instead of only unigrams, phrases ($n$-grams) are identified as potentially relevant descriptors of a document. Mihalcea et al. suggest that keyphrases can be constructed from keywords as post-processing step by collapsing co-occurring candidates into phrases \cite{mihalcea2004textrank}. The phrases are then ranked by averaging the scores of the individual terms contained in it. The primary task still remains efficiently extracting quality \textit{keywords} from the documents, which is why we focus on automatic keyword extraction problem.
  
%Words  in a document fall into two broad categories. Nouns, verbs, adjectives, adverbs, etc. are semantically important and are commonly accepted as \textit{content} words, while pronouns, articles, and prepositions are functors that support content words and are considered as \textit{non-content} words. Further, frequently used words, called {\it stopwords}, do not significantly impact the sense of the document, and are disregarded while performing automatic keyword extraction.

Earliest works on automatic keyword extraction employed purely statistical techniques based on term frequency to gauge importance of the words \cite{luhn1957statistical,sparck1972statistical}. Harter~\cite{harter1974probabilistic} and Bookstein et. al.~\cite{bookstein1974probabilistic} explored probabilistic approaches for automatic keyword indexing using 2-Poisson distribution model to represent \textit{specialty} words. According to another hypothesis, keywords follow a non-homogeneous distribution and tend to form clusters~\cite{ortuno2002keyword,zhou2003metric}. %These statistically inclined works did not explicitly discriminate between \textit{content} and \textit{non-content} words, though a manually curated contextual {\it stop-list} was used to omit some words from processing~\cite{bookstein1974probabilistic}. 
In recent years two lines of development of keyword extraction methods have gained prominence. First of these is the machine learning based approaches and the second is based on the graph representation of text.

{\it Machine learning approaches} come in supervised \cite{hulth2003improved, turney2000learning, witten1999kea} and unsupervised \cite{litvak2011degext, matsuo2001keyworld, mihalcea2004textrank} flavors. Supervised learning methods require labelled training data to induce the model. Each instance in the training set represents a term in  the document with  label 1 (keyword) or 0 (not a keyword). Creation of training set requires manual annotation of the text, making the task tedious, subjective, and possibly inconsistent. Because of the intense human intervention required,  supervised methods for keyword extraction have not been able to sustain interest and popularity. Due to this reason, unsupervised methods are favored as alternative approach for identifying keywords.%, since they do not require any labelled data to induce the model.

{\it Graph-based approaches} denote candidate keywords as nodes and the relationship between two nodes as an edge. Different types of scoring functions are used to rank the candidates based on specific graph property, e.g., centrality measure \cite{florescu2017position,lahiri2014keyword,litvak2011degext,mihalcea2004textrank}, k-degeneracy \cite{rousseau2015main,tixier2016graph}, etc. Performance of graph-based approaches is influenced by the pre-processing steps, graph construction method, and nature of the scoring function.

Existing state-of-the-art graph-based keyword extraction methods suffer from three limitations. First, the methods require user parameters during graph construction and word scoring stages \cite{mihalcea2004textrank,rousseau2015main,tixier2016graph}, which cast the burden of careful tuning of the parameters on the user. Second, the scoring methods  rely only on co-occurrence relation between the candidate keywords, while completely ignoring semantic relationship. Finally, these methods use linguistic tools to filter candidates from the document, limiting their use for many tool-poor languages. These observations  motivate - (i) design of \textit{parameterless graph-based method} for improving usability; (ii) design  of word scoring methods that account for  \textit{semantic connectivity} among the words, and (iii) development of \textit{language-independent} keyword extraction methods. Research in these directions is quintessential for advancing the state-of-the-art.

\subsection{Our contribution}
\label{subsec:contribution}
In this paper we present an in-depth study of current state-of-the-art graph-based keyword extraction methods. We advance the state-of-the-art by proposing two algorithms for automatic keyword extraction - one for languages with support of sophisticated NLP tools, and the other for languages that lack support of NLP tools, e.g., Indian languages. %Both keyword extraction methods are graph-based, parameterless , and profit from semantic connectivity among the words in a document. %Eliminating the need of NLP tools for candidate filtering, we design a language-agnostic keyword extraction algorithm to enable efficient and effective indexing of digital text documents of all languages. 
Specifically, our contributions are:

\begin{enumerate}
\item critical discussion of the issues and challenges of graph-based keyword extraction methods (Section \ref{sec:issuesGBKE}).
\item \label{item:CATG}design of a novel, parameterless method for constructing a context-aware graph of text (Section \ref{sec:CAGC}).%\par
\item \label{item:SCScore}design of a novel word scoring method that aims to capture (i) contextual hierarchy, (ii) semantic connectivity, and (iii) positional weight of the words in the text (Section \ref{sec:SCbased-word-score}).%\
\item experimental evaluation of items (ii) and (iii) individually, and comparison with counterparts in state-of-the-art methods (Sections \ref{subsec:compGraphConstruction} and \ref{subsec:compWordScore}).
\item design of a novel parameterless, \textit{s}emantic Connectivity Aware Keyword Extraction method (sCAKE) by  integrating (ii) and (iii), and its performance evaluation (Section \ref{sec:sCAKEvsPR}).
\item design of Language Agnostic Keyword Extraction method (LAKE) to extend keyword extraction service to languages that lack support of sophisticated NLP tools  (Section \ref{sec:lake-method}).
\end{enumerate}

%\subsection{Organization of the paper}
We review existing literature in Section \ref{sec:relwork}, followed by experimental setup and dataset details in Section \ref{sec:experimental-setup}. %Section \ref{sec:issuesGBKE} addresses the issues and challenges associated with graph-based keyword extraction methods. 
Please note that we are compelled to place experimental setup early in the paper because of our intention to investigate, both individually and together, the graph construction and word scoring methods in the state-of-the-art.  %Sections \ref{sec:CAGC} and \ref{sec:SCbased-word-score} propose novel graph construction and word scoring methods, respectively. Sections \ref{sec:sCAKEvsPR} and \ref{sec:lake-method} are devoted to discussion of sCAKE and LAKE algorithms, respectively. 
Section \ref{sec:conclusion} concludes the paper. We apologize for disappointing the reader who is looking for an explicit section on performance evaluation.

%%%%%%%%%%%%%%%%%%% SECTION: RELATED WORK %%%%%%%%%%%%%%%%%%%%
%\input{mycontent/Related-work}
\section{Related works}
\label{sec:relwork}

Works related to automatic keyword extraction methods emanate from largely four approaches. \textit{Statistics-based } approaches use simple and intuitive statistics like frequency \cite{luhn1957statistical,sparck1972statistical} and spatial distribution of terms \cite{bookstein1974probabilistic,harter1975probabilistic,herrera2008statistical,ortuno2002keyword,zhou2003metric} to identify candidate keywords. %The main strength of these approaches is their domain- and language-independence.  Sensitivity  to the spatial distribution of words in the document, however, is their main weakness.
\textit{Linguistic} approaches for identifying keywords use some form of linguistic analysis including lexical, semantic, and discourse analysis \cite{dostal2011automatic,ercan2007using,hulth2003improved,salton1991automatic}. % Non-availability of sophisticated NLP tools for most languages limits use of these methods for keyword extraction.  
\textit{Machine Learning} approaches (supervised and unsupervised) have found immense popularity in recent years, which involves training a model for identifying keywords from texts \cite{boudin2018unsupervised, frank1999domain, hulth2003improved, litvak2011degext, matsuo2001keyworld, mihalcea2004textrank, turney2000learning, witten1999kea, zhang2006keyword}. \textit{Graph-based} approaches represent the text as graph, where nodes denote unique terms and edges define the relationship among nodes. Candidate terms are ranked using either local or global graph properties \cite{florescu2017position, litvak2011degext, matsuo2001keyworld, mihalcea2004textrank, ohsawa1998keygraph, rousseau2015main}.% These algorithms minimize the impact of term frequency during the process of keyword extraction, while emphasizing co-occurrence of terms.

Since statistic- and  graph-based approaches are closely related to our work, we review selected research works from these areas in the following subsections.% with my-side bias towards the latter.

\subsection{Statistics-based Methods}

Statistical methods are the earliest keyword extraction techniques. The primary objective of early methods was to solve the problem of automatic indexing using term frequency \cite{luhn1957statistical,sparck1972statistical}. Luhn introduced Term Frequency (TF) to measure the extent of relevance of the words in a text document~\cite{luhn1957statistical}, which was later improved   by introducing Inverse Document Frequency (IDF) ~\cite{sparck1972statistical}. Words with high TF-IDF scores are considered important, and are used for indexing. One major limitation of TF-IDF method is its being corpus dependent, which restricts its applicability to dynamic collections. Later, Harter~\cite{harter1974probabilistic} and Bookstein et al.~\cite{bookstein1974probabilistic} explored the use of 2-Poisson distribution model to identify relevant terms in the document. Harter introduced a measure of \textit{indexability} to reflect the relative significance of words in a document~\cite{harter1975probabilistic}.

According to another hypothesis, keywords tend to exhibit high degree of self-attraction leading to non-homogeneous distribution that manifests as clusters \cite{ortuno2002keyword, zhou2003metric}. Ortuno et al. conjectured that the standard deviation of positions of occurrence of a word $w$ indicates its degree of relevance in the document, with higher values interpreted as higher degree of relevance~\cite{ortuno2002keyword}. Zhou and Slater advanced this idea and proposed two measures - $\sigma$-index and $\Gamma$-index to quantify relevance of words in text~\cite{zhou2003metric}. Computation of $\sigma$-index is similar to the approach proposed in \cite{ortuno2002keyword}, with minor modifications in the boundary conditions. Both $\Gamma$-index and $\sigma$-index exploit the spatial distribution of the words in the text document. Herrera et al.~\cite{herrera2008statistical} proposed an index for keyword extraction based on Shannon's entropy. Carratero et al. empirically showed that the entropy-based methods are sensitive to the choice of partition~\cite{carretero2013improving}, which is an undesirable property. %They proposed improvement over the clustering and entropy-based approaches by enhancing the boundary conditions, and concluded that clustering based approaches are comparatively better for short texts.

\subsection{Graph-based Methods}

%The variety of graph-based keyword extraction algorithms published since previous decade is the evidence of popularity of the approach
With words in the text represented as nodes, and relationship among them represented as edges, {\it graph of text} proved to be a rich and popular data model for analyzing text \cite{florescu2017position, litvak2011degext,matsuo2001keyworld, mihalcea2004textrank, ohsawa1998keygraph, rousseau2015main, tixier2016graph}. Blanco et al. describe different types of edge relationships that can be established among the nodes in a graph of text~\cite{blanco2012graph}. Term co-occurrence is the most commonly used relation, where the graph is constructed by linking the terms co-occurring within a window of pre-specified size. Subsequently, a word scoring mechanism that exploits discriminating properties of nodes is used to identify keywords.

KeyGraph method proposed by Ohsawa et al. segments the co-occurrence graph into clusters~\cite{ohsawa1998keygraph}, where each cluster corresponds to a concept. The terms in each cluster are ranked using a probability-based measure that quantifies the relationship of each term to the parent cluster, and top ranking terms are extracted as keywords. Mutsuo et al. established that co-occurrence text graphs exhibit `small-world' property~\cite{matsuo2001keyworld}. They proposed KeyWorld scoring method based on the contribution of each node of the graph to the small world property. TextRank~\cite{mihalcea2004textrank} is the most popular graph-based keyword extraction method so far. The method scores a node using PageRank \cite{brin1998anatomy} algorithm, which takes into account the global topology of the text graph. Litvak and Last proposed a degree based keyword extractor, DegExt, which exploits degree property of nodes and is computationally more efficient than TextRank~\cite{litvak2011degext}. PositionRank~\cite{florescu2017position} is an extension of TextRank that takes into account the positional information of terms in the document to assign weights to the candidate keywords, favoring words occurring towards the beginning of the text. This method reaffirms the positional importance of the words accorded by statistical methods.% by incorporating spatial information into computation of score.

Rousseau et al. hypothesized that the nodes participating in the most cohesive connected component of the text graph are apt candidates for keywords~\cite{rousseau2015main}. They performed core-based decomposition \cite{seidman1983network} of the graph to obtain the keywords. On a similar note, Tixier et al. \cite{tixier2016graph} performed truss-based decomposition \cite{cohen2008trusses} to retain words from the top-truss as keywords. These methods are parameter-free since the number of keywords extracted by these methods adapt to the structure of the graph.

\textit{We build over several of these ideas, including hierarchy used in \cite{rousseau2015main, tixier2016graph}, concepts in text \cite{ohsawa1998keygraph}, and importance of position of the word proposed in \cite{florescu2017position}, %Notably, we use hierarchy in an innovative manner to proxy for concepts and semantic connectivity  in the text.
and propose a parameterless keyword extraction algorithm.}% which takes into account the semantic connectivity of words in a document.}

%%%%%%%%%%%%%%%% EXPERIMENTAL SETUP %%%%%%%%%%%%%%%%%%%%
%\input{mycontent/Experimental-setup}
\section{Experimental Setup}
\label{sec:experimental-setup}
We use R (version 3.3.1) and Python (version 2.7.12) for implementation\footnote{The code for implementation is available at \url{https://github.com/SDuari/sCAKE-and-LAKE}}, using functions from NLP, igraph, openNLP, tm, foreach, and doSNOW packages\footnote{\url{https://cran.r-project.org/web/packages/}}. We execute the programs on a 64-bit PC with 8GB RAM, and Intel Core i7-6700 CPU @ 3.40GHz 8-core Processor running Ubuntu 16.04 LTS.

We use four benchmark datasets shown in Table \ref{tab:datasets} for empirical observations and comparisons. These datasets have been used extensively to evaluate keyword extraction algorithms \cite{boudin2013comparison, hulth2003improved, mihalcea2004textrank, rousseau2015main, savova2010mayo, you2013automatic}. Table \ref{tab:datasets} presents general properties of the four datasets, including number of documents in corpus, average document length, average number of gold-standard keywords along with standard deviation, and average percentage of candidate keywords. Hulth2003 documents, which are abstracts, are the shortest. Krapivin2009 documents have least average number of keywords assigned to them. It is noteworthy that the average number of candidates lies in the range of 40-45$\%$ of the document length.% for each dataset.
\begin{table}[ht!]
\scriptsize
\caption{\label{tab:datasets}Overview of experimental datasets. $|D|$: Number of documents in corpus, L: average document length, Avg/sd: average number of gold-standard keywords per document/standard deviation, C: average percentage of candidate keywords (nouns and adjectives)}
\begin{minipage}{\columnwidth}
\begin{center}
\begin{threeparttable}
\begin{tabular}{ c c c c c c }
\hline
\textbf{Dataset} & $\mathbf{|D|}$ & \textbf{L} & \textbf{Avg/sd} & \textbf{C} & \textbf{Dataset Description}\\ 
\hline
\makecell{Hulth2003\tnote{*} \cite{hulth2003improved}} & 1500 & 129 & 23/12.44 & 45.97 & Abstracts from \textit{Inspec} database\\
\makecell{NLM500 \cite{aronson2000nlm}} & 500 & 4854 & 27/10.38 & 44.08 & Full papers from \textit{PubMed} database\\
\makecell{Krapivin2009 \cite{krapivin2009large}} & 2304 & 7961 & 11/6.44 & 40.5 & ACM full papers\\
\makecell{SemEval2010\tnote{*} \cite{kim2010semeval}} & 244 & 8085 & 34/10.35 & 40.05 & ACM Digital Library papers\\
\hline
\end{tabular}
 \begin{tablenotes}
  \item[*] We use Test and Training Sets.
 \end{tablenotes}
\end{threeparttable}
\end{center}
%\bigskip
\end{minipage}
\end{table}

For evaluation, we use the uncontrolled list of keywords for Hulth2003, gold-standard keywords for Krapivin2009 and NLM500,  and author-and-reader-assigned keywords for Semeval2010. We use classical F1-measure to evaluate performance of the compared algorithms for top-$k$ extracted keywords. The results are macro-averaged at the dataset level. We consider TextRank \cite{mihalcea2004textrank}, DegExt \cite{litvak2011degext}, $k$-core retention \cite{rousseau2015main}, and PositionRank \cite{florescu2017position} as our prime competitors and evaluate the proposed approaches against them. 

For each dataset, we experimented with all algorithms to find the value of $k$ that yields the best F1-measure. It was observed that the highest F1-measure was obtained for $k = 25$ for Hulth2003, $k = 10$ for Krapivin2009, and $k = 30$ for NLM500 and SemEval2010 datasets. We use these values of $k$ for reporting results for corresponding datasets in subsequent experiments. It is pertinent to note that the values correlate with the average number of gold-standard unigrams (Column 4 of Table \ref{tab:datasets}) annotated for the datasets.

%%%%%%%%%%%%%%%% GBKE ISSUES AND CHALLENGES %%%%%%%%%%%%%%%%%%%%
%\input{mycontent/GBKE-issues-and-challenges}
\section{Graph-based Keyword Extraction: Issues and Challenges}
\label{sec:issuesGBKE}
\begin{figure}[hbtp]
\centering
\begin{minipage}{.75\linewidth}
\includegraphics[scale=0.33]{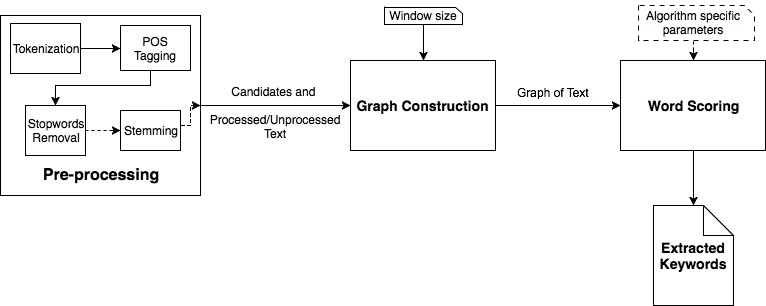}
\end{minipage}%
%\hspace{4.5cm}
\begin{minipage}{.25\linewidth}
\caption{Sequence of sub-tasks in graph-based keyword extraction methods}
\label{Fig:GBKE-flow}
\end{minipage}
\end{figure}

Graph-based keyword extraction algorithms perform three generic steps in sequence - (i) pre-processing of text to identify candidate keywords, (ii) transforming text to graph with candidates as nodes, and (iii) scoring the candidates based on some local or global graph property. Figure \ref{Fig:GBKE-flow} depicts the process of graph-based automatic keyword extraction. It is the variation in design of the core steps and their execution that produces a bouquet of graph-based keyword extraction algorithms \cite{erkan2004lexrank, florescu2017position, grineva2009extracting, litvak2011degext, liu2010automatic, matsuo2001keyworld, mihalcea2004textrank, ohsawa1998keygraph, rousseau2015main}. In the following subsections, we discuss the variations of these three steps and deliberate on the issues and challenges faced by graph-based keyword extraction approaches. Each subsection focuses on one task, delineates the challenges, and describes how the challenges are addressed by the existing algorithms. We support our arguments with empirical evidences, wherever relevant.
\subsection{Pre-processing of Text}
\label{subsec:preprocessing}
Pre-processing of text significantly affects the resulting keywords because the output from this step is the primary input to the graph construction phase. A different combination of pre-processing sub-steps has a defining effect on performance of the methods. Tokenization and stopword\footnote{Frequently used words, called {\it stopwords}, are disregarded during automatic keyword extraction.} removal are performed by all algorithms~\cite{florescu2017position, litvak2011degext, mihalcea2004textrank, rousseau2015main}. Barring DegExt, all algorithms perform POS tagging and agree that nouns and adjectives are the prime candidates for keywords~\cite{florescu2017position, mihalcea2004textrank, rousseau2015main}.  DegExt doesn't inflict any restriction over the candidates except for stopwords, which are similarly disregarded in all methods. Only $k$-core retention algorithm~\cite{rousseau2015main} uses stemming, and claims that it boosts performance.

Average recall for any algorithm for a particular document is bounded by the percentage of gold-standard keywords actually present in the document. We studied the gold-standard keyword lists of the four datasets and found that stemming increases the upper bound for recall in all datasets. First column in Table \ref{tab:stem_effect} shows this bound without stemming the documents, and the second column shows the bound after stemming. 
\begin{table}[ht]
\begin{minipage}{.6\linewidth}
\scriptsize
\begin{center}
\begin{tabular}{c c c}% m{1.3cm} | m{2cm} | }
\hline
\textbf{Dataset} & \textbf{w/o stemming} & \textbf{With stemming} \\
\hline
Hulth2003 & 89.86013 & 92.0831\\
NLM500 & 70.58481 & 79.2508\\
%\hline
Krapivin2009 & 96.88258 & 98.17081\\
%\hline
SemEval2010 & 95.91513 & 98.95135\\
\hline
\end{tabular}
\end{center}
\end{minipage}
%\hspace{2.5cm}
\begin{minipage}{.4\linewidth}
\caption{Percentage of gold-standard keywords present in text with and without stemming\label{tab:stem_effect}}
\end{minipage}
\end{table}

\textit{The issue of effective sequence of pre-processing steps for keyword extraction is more or less settled. However, a vast majority of languages fail to benefit from existing keyword extraction methods due to the lack of sophisticated NLP tools required for pre-processing by these methods. We address this issue later in Section \ref{sec:lake-method}.}

\subsection{Graph Construction}
\label{subsec:graph_construction}
Existing keyword extraction algorithms exhibit wide variations in the process of constructing graph from text. The resulting structural differences naturally cascade into differential in graph properties. Since graphs are principal inputs for ranking the candidates, the word scores and the set of extracted keywords veritably differs for different algorithms.

Variations in graph construction methods align primarily in two dimensions. First, the set of candidate keywords obtained after pre-processing the text. This impacts the order\footnote{Order of a graph is the cardinality of the node set.} of the graph and its properties. For example, candidate lists produced after stemming creates a smaller graph as compared to those produced without stemming. Second is the scheme for defining relationship between the nodes (i.e. the edge set), which affects the construction and size\footnote{Size of a graph is the cardinality of the edge set.} of text graph. Edge direction and edge weight are other considerations for graph construction. DegExt \cite{litvak2011degext} constructs unweighted, directed graph corresponding to the order of words in original text. Other methods construct weighted, undirected graph of text where edge weight is the frequency of co-occurrence of the two words.

Variations in the text graphs are more conspicuous because of the second dimension. Two parameters, viz. window-size and source text for sliding the window emerge as fundamental causes of differences in edge sets and the resultant graphs. Though all existing algorithms use co-occurrence of words within a specified \textit{window} as the relationship, it is the size of the window that induces pronounced differences. Different keyword extraction methods recommend different window sizes. TextRank suggests window size of 2-10 and compares 2, 3, 5, and 10 for experimental evaluation (Page 5, Table 1 of \cite{mihalcea2004textrank}). DegExt uses window of size 2 that does not connect words separated by punctuation marks (Page 3, \cite{litvak2011degext}), while $k$-core retention algorithm uses window of size 4 (Page 4, \cite{rousseau2015main}). Apparently, the choice of window size parameter in all works is based on empirical observation over the experimented datasets.  

Differences in the text graphs are further accentuated by the source text where the relationship is examined. Some methods recommend sliding the window on raw text \cite{florescu2017position, mihalcea2004textrank}, while others slide on  pre-processed text \cite{litvak2011degext, rousseau2015main}.  There is no systematic and scientific study of these two parameters (window size and source text) of graph construction methods to the best of authors' knowledge. Lack of consensus on these two issues poses difficult decision choices for the users and the designers of the algorithm.
\begin{table}[ht]
\scriptsize
\begin{minipage}{4cm}
%\caption{Comparison of graph-construction methods.\label{tab:graphconstruction} $w$: window size parameter, Source: text to slide window, Overspan: whether to connect words separated by punctuation marks or not?, UD:Undirected, D: Directed.}
\begin{center}
\begin{tabular}{ c c c c c c }
\hline
\multirow{2}{*}{\textbf{Graph}}  & \multicolumn{5}{c}{\textbf{Graph construction}}\\
\cline{2-6}
& $w$ & Directed & Weighted? & Source & Overspan\\
\hline
TG  & 2 to 10 & No & Yes & Original & Yes\\
GoW & 4 & No & Yes & Processed & Yes\\ 
DG & 2 & Yes & No & Processed & No\\ 
%PositionRank & TG & 2 to 10 & UD & Yes & Original & Yes\\
\hline
\end{tabular}
\end{center}
\end{minipage}
\hspace{4cm}
\begin{minipage}{4cm}
\caption{Comparison of graph-construction methods.\label{tab:graphconstruction} $w$: window size parameter, Source: text to slide window, Overspan: connect words separated by punctuation marks.}
\end{minipage}
\end{table}

Table \ref{tab:graphconstruction} summarizes the differences in graph construction approaches adopted by the state-of-the-art keyword extraction methods. In this table (and all others following), we use acronyms for graphs created by TextRank and PositionRank\footnote{PositionRank uses same settings as TextRank for pre-processing and graph construction.} (TG), DegExt (DG), and $k$-core retention (GoW) algorithms. Figure \ref{Fig:graphs} shows the graphs constructed by three different algorithms (\ref{Fig:core-g}, \ref{Fig:tr-g}, and \ref{Fig:deg-g}) for the same sample text (\ref{Fig:text}), highlighting the  differences among the graph construction approaches.

\begin{figure*}[h!]
\begin{subfigure}{\columnwidth}
\centering
\includegraphics[scale=0.45]{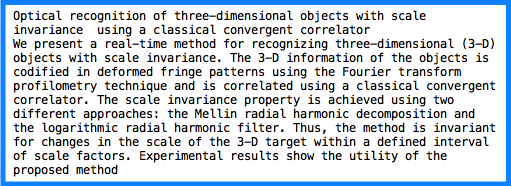}
\caption{Text document}
\label{Fig:text}
\end{subfigure}
\begin{subfigure}{\columnwidth}
\centering
\includegraphics[scale=0.5]{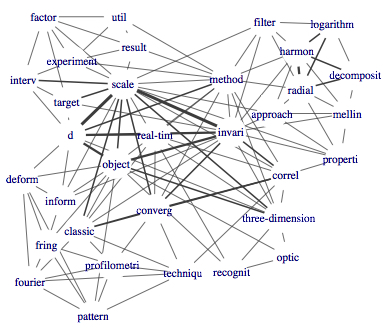}
\caption{Graph-of-Word (GoW) \cite{rousseau2015main}}
\label{Fig:core-g}
\end{subfigure}
\begin{subfigure}{\columnwidth}
\centering
\includegraphics[scale=0.5]{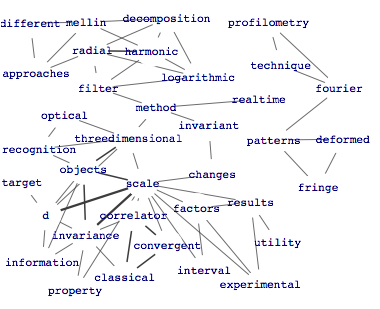}
\caption{TextRank Graph (TG) \cite{florescu2017position, mihalcea2004textrank}.}
\label{Fig:tr-g}
\end{subfigure}
\end{figure*}
\clearpage
\begin{figure*}[ht!]
\ContinuedFloat
\begin{subfigure}{\columnwidth}
\centering
\includegraphics[scale=0.5]{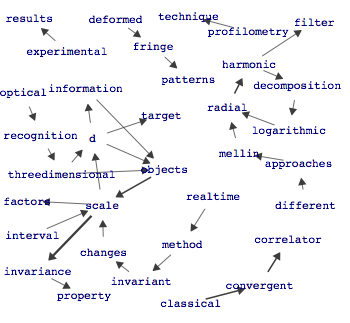}
\caption{DegExt Graph (DG) \cite{litvak2011degext}}
\label{Fig:deg-g}
\end{subfigure}
\caption{Text graphs created by different algorithms for document id 2015 of Hulth2003 Test dataset. Edge width is proportional to the corresponding edge weight. TG and GoW graphs are constructed with window-size 4 and DG graph with window-size 2.}
\label{Fig:graphs}
\end{figure*}

\subsection{Word Scoring}
\label{subsec:keyword_scoring}
Word scoring methods are crucial discriminators between keyword extraction algorithms. TextRank \cite{mihalcea2004textrank} uses PageRank \cite{brin1998anatomy} algorithm to assign importance to candidates by recursively taking into account importance of its neighbors. % its relationship (co-occurrence) with its neighbours.
Thus, the knowledge drawn from the global graph structure is used to rank the words. TextRank uses a parameter called damping factor\footnote{Damping factor is associated with the concept of {\it random jump} in web search.} $d$, which is set to 0.85 following~\cite{brin1998anatomy}. We examine the impact of damping factor on performance of the algorithm. Our experiments on Hulth2003 dataset (used for evaluation by TextRank) reveal that best performance is achieved for different values of damping factor for different window sizes. Specifically, the best result in terms of F1-score is obtained for window-sizes 2, 3, and 4 when $d$ is set to 0.85, 0.9, and 0.95, respectively. Among different combinations of the two parameters, window-size 4 and $d=0.95$ yields best result. This is purely an empirical observation specifically for this dataset. We are not in position to ascribe any theoretical reason to the phenomenon, but state this to highlight the sensitivity of the end results towards the algorithmic parameters. We use window-size $4$ and $d=0.95$ for subsequent experiments in accordance with our observation.

DegExt~\cite{litvak2011degext} uses degree centrality to score the relevance of candidates. Authors claim to achieve performance comparable to TextRank with lesser computational complexity. K-core retention algorithm~\cite{rousseau2015main} doesn't score candidates explicitly. Instead, it uses core decomposition of the weighted graph and retains words from the top core as keywords. %DegExt and K-core retention algorithm are parameter-free for their respective word scoring approaches. 
PositionRank~\cite{florescu2017position} uses position-biased PageRank to rank the candidates by favoring words that occur towards the beginning of the text, and %Recall that the position of the word in the text is considered discriminative by some statistical approaches \cite{ortuno2002keyword, zhang2008automatic, zhou2003metric}. 
uses same parameters as TextRank. Table \ref{tab:wordscoring} summarizes the word-scoring methods and their respective parameters for four methods.
\begin{table}[!hbtp]
\scriptsize
\begin{minipage}{.6\linewidth}
\begin{center}
\begin{tabular}{ c c c c }
%\toprule
\hline
\multirow{2}{*}{\textbf{\makecell{KE\\Algorithms}}} & \multicolumn{3}{c}{\textbf{Word Scoring}}\\
\cline{2-4}
%\cline{7-9}
& \textbf{\makecell{Scoring Method}} & $\mathbf{S_{params}}$ & \textbf{Value}\\
%\midrule
\hline
\multirow{3}{*}{TextRank} & \multirow{3}{*}{PageRank} & $d$ & $ 0.85$\\
& & $t$ & 1e-4\\ 
& & $n$ & Top 1/3\\ 
\hline
K-core & \makecell{Weighted $k$-core\\decomposition} & - & -\\ 
\hline
DegExt & \makecell{Degree Centrality} & $n$ & U\\ 
\hline
\multirow{3}{*}{PositionRank} &\multirow{3}{*}{\makecell{Position-biased \\PageRank}} & $d$ & $ 0.85$\\
& & $t$ & 1e-3\\ 
& & $n$ & U\\
%\bottomrule
\hline
\end{tabular}
\end{center}
\end{minipage}
%\hspace{4cm}
\begin{minipage}{.4\linewidth}
\caption{Comparison of word scoring methods for different Keyword Extraction algorithms\label{tab:wordscoring}. $S_{params}$: Algorithmic parameters for word scoring method as used by published works, $d$: damping factor, $t$: convergence threshold, $n$: Number of keywords to be extracted, U: User parameter. }
\end{minipage}
%\end{center}
\end{table}

{\it To the best of authors' knowledge, investigation of the combination of pre-processing and graph construction methods that yields best performance for keyword extraction methods is pending.}

\subsection{Evaluation of Keyword Extraction Methods}
\label{subsec:kem-issues}
No system is capable of definitive assessment of relevance of the words in a document because relevance is subjective not only with respect to the reader of the document, but also with respect to time. Further, evaluation of keyword extraction method is based on the assumption that importance of words is a dichotomous variable that is user-specific, and is defined outside the system. It is therefore imperative to evaluate keyword extraction methods against a gold-standard keywords list.

Automatic keyword extractors are judged on the basis of how \textit{precisely} they extract and how well they {\it recall} the  keywords that exist in gold-standard keywords list \cite{florescu2017position, hulth2003improved, litvak2011degext, mihalcea2004textrank, rousseau2015main,  witten1999kea}. Gold-standard keywords are manually annotated, and hence often subjective and noisy. Over- and under-annotation in gold-standard lists influence the performance of keyword extractors. Consequently, performance of one algorithm may be different for different datasets. Recently, Florescu et al. \cite{florescu2017position} used mean reciprocal rank (MRR) to evaluate the performance of their algorithm, which is based on the single highest-ranked relevant item. However, we believe that MRR is better suited for evaluation of web search methods where the single highest-ranked relevant item is important for the user. Since number of keywords required is more than one, MRR could be misleading for evaluating performance of automatic keyword extractors.

%Interestingly, the metrics for performance evaluation have largely remained unchanged over several years. Precision, recall, and F-measure are used with occasional modifications to the original design \cite{florescu2017position, hulth2003improved, litvak2011degext, mihalcea2004textrank, rousseau2015main,  witten1999kea}. 

%The number of keywords to be extracted is a crucial parameter for performance evaluation of automatic keyword extraction methods. 
Most algorithms accept the number of keywords to be extracted as a user parameter~\cite{florescu2017position, litvak2011degext, turney2000learning} or a pre-decided value \cite{mihalcea2004textrank}. Alternatively, this number can be set to take a value based on the structure of the text graph~\cite{rousseau2015main, tixier2016graph}. Higher value of this parameter is often associated with higher recall and low precision, while lower value is associated with lower recall and high precision \cite{manning2008introduction}. Algorithms may choose to match either \textit{unigrams} \cite{rousseau2015main, tixier2016graph} or \textit{keyphrases} \cite{florescu2017position, mihalcea2004textrank} against the gold standard list. However, there is no consensus in literature regarding evaluation of \textit{keyphrase extraction approaches}, as it is not clear whether to reward or penalize a method that over- or under-estimates keyphrases given the gold-standard list \cite{rousseau2015main}. 

\textit{Performance of keyword extraction methods varies depending on the parameter settings used, as well as the properties of experimental datasets. No algorithm is able to perform uniformly well across domains and corpora.}

%%%%%%%%%%%%%%%% CAG GRAPH CONSTRUCTION %%%%%%%%%%%%%%%%%%%%
%\input{mycontent/graph-constr}
\section{Context-aware Graph Construction Method}
\label{sec:CAGC}
Motivated by the desideratum to design parameterless graph construction method, we propose to construct co-occurrence graph based on pragmatics of written communication. Unlike semantics, which studies the meaning coded in the language, pragmatics involves study of transmission of meaning depending on the context of utterance. The context set by a sentence is often used by the consecutive sentences, imparting continuity in communication. This phenomenon, called entailment, is a well studied concept in linguistics. Transmission of context from one sentence to another is the core idea underlying the proposed co-occurrence graph construction method. 

In this method, the window slides over two consecutive sentences and the candidates co-occurring therein are linked. This eliminates the need of integer-valued \textit{window-size} parameter,  %This makes graph construction  parameter-free by getting rid of the window-size parameter used by the state-of-the-art methods. 
and captures contextual co-occurrence of words\footnote{We use `word', `node' and `term' interchangeably in the rest of the paper.} (terms) in text. The resulting graph, called Context-Aware Text Graph (CAG), is formally represented as $G_{CAG} = (V,E,W)$. Here, $V$ is the set of nodes representing the candidate words, $E$ is the set of edges (co-occurrence relation), and $W$ is the set of corresponding edge weights. Weight $w_{ij}$ for an edge $e_{ij}$ indicates the co-occurrence frequency of two words $v_i$ and $v_j$ in the text. Higher value of $w_{ij}$ indicates stronger contextual relationship between words $v_i$ and $v_j$.

For graph creation, we consider two consecutive sentences ($s_k$ and $s_{k+1}$) in the given text as one document ($d_k$), and create a Boolean term-document matrix $C$, where 
\[
   c_{ik}= 
\begin{cases}
  1, & \text{if term } t_i \text{ occurs in } d_k\\
  0, & \text{otherwise}
\end{cases}
\]

\begin{figure}[ht!]
\centering
\begin{minipage}{.7\linewidth}
    \includegraphics[scale = 0.5]{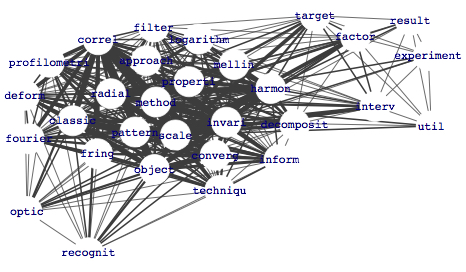}
\end{minipage}%
%\hspace{4cm}
\begin{minipage}{.3\linewidth}
    \caption{Context-Aware Text Graph for the sample text shown in Figure \ref{Fig:text}.}
\label{Fig:graph-ldke}
\end{minipage}
\end{figure}

In accordance with the convention, we use the set of nouns and adjectives as candidates to construct matrix $C$. Let $T = CC^\mathsf{T}$ denote the term-term matrix where $\tau_{ij}$ represents the number of co-occurrences of terms $t_i$ and $t_j$ in the documents (pairs of consecutive sentences). %($d_k, k \in \{1, 2, ..., K-1\}$, where $K$ is the number of sentences).
Note that $T$ is the symmetric adjacency matrix of an undirected, weighted graph $G$. The context-aware text graph, $G_{CAG}(V,E,W)$, is constructed from $T$ after zeroing the diagonal elements. Figure \ref{Fig:graph-ldke} shows a sample graph created using the proposed CAG method for the text shown in Figure \ref{Fig:text}. We observe that the graph created by CAG method is denser than those in Figure \ref{Fig:graphs}. This is because of the bigger co-occurrence span (two consecutive sentences) used in CAG method.% that connects more words with each other, resulting in higher number of edges.
\subsection{Comparison of Graph Properties}
\label{subsec:compGraphProperties}
We analyze the structural properties of the TG, DG, GoW, and CAG graphs for the four datasets mentioned in Section \ref{sec:experimental-setup}. We construct four types of graphs for each document in the datasets, and compute number of nodes and edges, global clustering coefficient\footnote{Also called transitivity of graph $G$.} \cite[p~101]{zaki2014data}, average path length \cite[p~98]{zaki2014data}, and density \cite[p~101]{wasserman1994social}. Tables (\ref{tab:diff-topology-hulth}-\ref{tab:diff-topology-sem}) show the variations in topological properties of graphs by averaging the results at the dataset level.
\begin{table}[!htb]
\setlength\tabcolsep{3pt}
    \scriptsize
    \caption{\label{tab:diff-topology}Topological properties of graphs constructed using TextRank (TG), DegExt (DG), $k$-core (GoW), and CAG for four dataset. $|V|$: number of nodes, $|E|$: number of edges, $CC$: global clustering coefficient, $APL$: average path length, $\Delta$: Density}
    
    \begin{subtable}{.5\linewidth}
      \centering
      %\begin{minipage}{3cm}
        \caption{\label{tab:diff-topology-hulth}Hulth2003 dataset.}
        \begin{tabular}{ c c c c c c }
			    \hline
			    \textbf{Method} & $\mathbf{|V|}$ & $\mathbf{|E|}$ & $\mathbf{CC}$ & $\mathbf{APL}$ & $\mathbf{\Delta}$ \\
			    \hline
			    TG & 39 & 67 & 0.40 & 3.37 & 0.11\\ %\hline
			    DG & 37 & 37 & 0.05 & 3.91 & 0.033\\ %\hline
			    GoW & 35 & 143 & 0.49 & 2.10 & 0.27\\ %\hline
			    CAG & 33 & 370 & 0.85 & 1.30 & 0.70\\ %\bottomrule
		    	\hline
		\end{tabular}
	  %\end{minipage}
    \end{subtable}%
    %\hspace{.3cm}
    \begin{subtable}{.5\linewidth}
      \centering
        \caption{\label{tab:diff-topology-krapi}Krapivin2009 dataset.}
        \begin{tabular}{ c c c c c c }
        	    %\toprule
        	    \hline
        	    \textbf{Method} & $\mathbf{|V|}$ & $\mathbf{|E|}$ & $\mathbf{CC}$ & $\mathbf{APL}$ & $\mathbf{\Delta}$ \\
            	%\midrule
        	    \hline
        	    TG & 716 & 2930 & 0.15 & 3.32 & 0.012 \\ %\hline
        	    DG & 697 & 1636 & 0.078 & 5.11 & 0.004\\ %\hline
        	    GoW & 555 & 5022 & 0.21 & 2.57 & 0.035\\ %\hline
        	    CAG  & 471 & 19664 & 0.51 & 1.87 & 0.16\\ %\bottomrule
        	    \hline
        \end{tabular}
    \end{subtable}
    %\vspace{.5cm}
    \begin{subtable}{.5\linewidth}
      \centering
        \caption{\label{tab:diff-topology-nlm}NLM500 dataset.}
        \begin{tabular}{ c c c c c c }
        	    %\toprule
        	    \hline
        	    \textbf{Method} & $\mathbf{|V|}$ & $\mathbf{|E|}$ & $\mathbf{CC}$ & $\mathbf{APL}$ & $\mathbf{\Delta}$ \\
        	    %\midrule
        	    \hline
        	    TG & 589 & 2083 & 0.17 & 3.62 & 0.013\\ %\hline
        	    DG & 540 & 938 & 0.06 & 6.04 & 0.004\\ %\hline
        	    GoW & 479 & 3647 & 0.22 & 2.67 & 0.036\\ %\hline
        	    CAG & 397 & 11514 & 0.44 & 1.90 & 0.151\\ %\bottomrule
        	    \hline
        \end{tabular}
    \end{subtable}%
    %\hspace{.5cm}
    \begin{subtable}{.5\linewidth}
      \centering
        \caption{\label{tab:diff-topology-sem}Semeval2010 dataset.}
        \begin{tabular}{ c c c c c c }
        	    %\toprule
        	    \hline
        	    \textbf{Method} & $\mathbf{|V|}$ & $\mathbf{|E|}$ & $\mathbf{CC}$ & $\mathbf{APL}$ & $\mathbf{\Delta}$ \\
        	    %\bottomrule
        	    \hline
        	    TG & 770 & 3085 & 0.15 & 3.35 & 0.012\\ %\hline
        	    DG & 727 & 1528 & 0.071 & 5.32 & 0.003\\ %\hline
        	    GoW & 617 & 5385 & 0.20 & 2.63 & 0.029\\ %\hline
        	    CAG & 507 & 13441 & 0.38 & 1.99 & 0.105\\ %\bottomrule
        	    \hline
        \end{tabular}
    \end{subtable}
\end{table}

Even though the four algorithms consider nouns and adjectives as candidates, average number of vertices in each graph type differs depending on the nature of edge connections. The variation in edge relation resulting due to different window size yields distinct sets of isolated vertices, which when excluded from the graphs results in different node sets. Some observations about CAG graphs are - (i) number of nodes is minimum in CAG method for all datasets, (ii) number of edges is highest for CAG because the co-occurrence span is usually larger than the window sizes adopted in other methods, and (iii) CAG graphs are denser than other graphs. The number of edges created per window slide depends on the number of candidates present within the co-occurrence span. Maximum number of edges created each time the integer-valued window of size $w$ slides is $(w-1)$, whereas for CAG it is bounded by $(|S_i|+|S_{i+1}|-1)$, where $|S|$ is the number of words in the sentence. This makes the CAG graph denser than the other algorithms. 

Due to the dense nature of CAG graphs, clustering coefficient is higher and average path length is lower for CAG. Other graphs are visibly less dense as compared to CAG graphs (Figures \ref{Fig:graphs} and \ref{Fig:graph-ldke}). DG graphs are the most sparse among these four types and thus have lowest clustering coefficient and highest average path length. This is due to the fact that DG uses a window-size of 2 as co-occurrence span for connecting nodes, which results in nodes being connected to a fewer nodes than the other three methods. {\it Variations in the structural properties of graphs play an instrumental role in word scoring, as discussed in the following subsection.}

\subsection{Performance Evaluation of CAG}
\label{subsec:compGraphConstruction}
We compare effectiveness of the four state-of-the-art graph construction methods with the proposed CAG method by applying native word scoring methods on the respective graphs, as well as on the context-aware graphs. Following the approach adopted by Rousseau et al. \cite{rousseau2015main}, we match $k$ keywords (as defined in Section \ref{sec:experimental-setup}) extracted from each document against the gold-standard keywords (as unigrams) to compute the performance evaluation metrics. Table \ref{tab:perf-eval-graphs} shows the experimental results as macro-averaged F1-score. 
\begin{table}[ht]
\scriptsize
       \caption{Comparative Evaluation of original vs. CAG graphs for  native scoring methods in terms of macro-averaged F1-score.\label{tab:perf-eval-graphs} PageRank, Degree, $k$-core, bised PageRank: word scoring methods for TextRank, DegExt, K-core, and PositionRank respectively}
       \begin{minipage}{\columnwidth}
\begin{center}
        \begin{tabular}{ c c c c c c }
        	%\toprule
        	\hline
        	\multirow{2}{*}{\textbf{\makecell{Word Scoring\\Methods}}} & \multirow{2}{*}{\textbf{Graph}} & \multicolumn{4}{c}{\textbf{Datasets}}\\
        	\cline{3-6}
      		& & \textbf{Hulth2003} & \textbf{Krapivin2009} & \textbf{NLM500} & \textbf{Semeval2010}\\
        	\hline
        	\multirow{2}{*}{PageRank} & Original & 18.37 & 13.72 & 10.73 & 13.65\\
        	 & CAG & 49.54 & 35.05 & 25.68 & \textbf{41.54} \\ \hline
        	\multirow{2}{*}{Degree} & Original & 18.22 & 13.34 & 10.91 & 14.36 \\
        	& CAG & 49.42 & 34.92 & 25.59 & 40.81 \\ \hline
        	\multirow{2}{*}{$k$-core} & Original & 43.41 & 22.70 & 20.20 & 29.34\\
        	 & CAG & 34.84 & 3.46 & 2.12 & 3.60 \\ \hline
        	\multirow{2}{*}{biased PageRank} & Original & 50.41 & 37.07 & 21.94 & 27.50 \\
        	& CAG & \textbf{51.01} & \textbf{42.86} & \textbf{27.54} & 35.80 \\ %\hline
        	\hline
        \end{tabular}
   \end{center}
\end{minipage}
\end{table} 

We observe that CAG graphs significantly boost F1-score of all scoring methods except $k$-core retention. Applying $k$-core decomposition on CAG graphs results in fewer nodes at the top core. %(1 or 2 nodes in most cases), because less number of high-degree nodes co-occur with other similarly high-degree nodes. 
This decreases recall significantly even though the precision is high, leading to a drastic drop in F1-score. We also note that PositionRank outperforms TextRank, K-core retention, and DegExt when applied on CAG graphs. This experiment establishes the effectiveness of context-aware graph construction method.% which can be used in conjunction with any word scoring method to boost performance.
\textit{This also affirms that capturing the context in the window that spans two consecutive sentences highlights the important words irrespective of the scoring method used.}

\subsection{Timing Comparison for Graph Construction}
\label{subsec:time_comp}
In order to gauge the computational efficiency, we compare the time taken by the four graph construction methods (including pre-processing). Table \ref{tab:time_comp} shows average time required per document to construct text graphs of four datasets. The timings (in seconds) are averaged over three executions for each data set. 
\begin{table}[ht]
\scriptsize
       \caption{Average time (in seconds) taken per document by the four algorithms on each dataset. Best timings are presented in bold.\label{tab:time_comp}}
       \begin{minipage}{\columnwidth}
\begin{center}
        \begin{tabular}{ c c c c c }
        	%\toprule
        	\hline
        	\multirow{2}{*}{\textbf{Methods}} & \multicolumn{4}{c}{\textbf{Datasets}} \\
        	\cline{2-5}
      		& \textbf{Hulth2003} & \textbf{Krapivin2009} & \textbf{NLM500} & \textbf{Semeval2010}\\
        	%\midrule
        	\hline
        	TG & 0.5245 & 35.30 & 30.94 & 52.84 \\ %\hline
        	DG & 0.3937 & 20.88 & 8.022 & 12.78 \\ %\hline
        	GoW & 0.0859 & 16.88 & 20.97 & 43.71 \\ %\hline
        	CAG  & \textbf{0.079} & \textbf{3.080} & \textbf{1.895} & \textbf{3.412} \\ %\bottomrule
        	\hline
        \end{tabular}
   \end{center}
\end{minipage}
\end{table} 

CAG method is found to execute significantly faster than other three methods on all datasets. It is important to note that the number of sentences is much less than the number of distinct windows of size $w$. For a document of length $N$ consisting of $S$ sentences ($N\gg S$), the co-occurrence identification in sliding window based algorithms is processed $(N-w)$ times, while in CAG it is processed $(S-1)$ times. This explains the speedy execution of CAG method.

%%%%%%%%%%%%%%%% SCScore WORD SCORING %%%%%%%%%%%%%%%%%%%%
%\input{mycontent/SCbased-Word-scoring}
\section{Semantic Connectivity based Word Scoring Method}
\label{sec:SCbased-word-score}
We exploit semantic connectivity between words in a document to identify important and relevant words, and propose a novel word scoring method. The proposed method leverages - (i) the level of hierarchy of a word in text graph, (ii) its semantic relationship with neighbors, (iii) the extent of its semantic connectivity, and (iv) its positions of occurrence in the text. It is pertinent to note that we do not use any linguistic tool to capture semantic aspects.

\subsection{Level of Hierarchy}
Recently, it has been established that hierarchy of nodes (words) in co-occurrence graphs is the sole determinant of the importance of the word~\cite{rousseau2015main, tixier2016graph}. Rousseau et al. \cite{rousseau2015main} used core-based decomposition and Tixier et al. \cite{tixier2016graph} used truss-based decomposition to obtain the hierarchy. %According to this view, the words in the top of hierarchy of core-based decomposition of the graph are the keywords.
Though efficient, these methods have two major limitations. First, not only the keywords but even the number of keywords in the text is determined singularly by the hierarchy. This may result in too few keywords, thereby degrading the recall. Second, both decomposition methods have low semantic interpretability when used singly.

Subscribing to the view advanced by Tixier et al., which shows truss-based decomposition works better than core-based decomposition, we use trussness~\cite{cohen2008trusses} to elicit hierarchy of words in the graph. Truss-based decomposition is a graph peeling algorithm that results into a sequence of subgraphs (called trusses), each of which is denser than the previous one. Cohen observed and we quote ``The $k$-truss provides a nice compromise between the too-promiscuous $(k-1)$-core and the too-strict clique of order $k$"~\cite{cohen2008trusses}. We briefly introduce $k$-truss and the concept of trussness below, adapting definition from \cite{cohen2008trusses}.

\begin{defn}
For a weighted, undirected, simple graph $G = (V,E,W)$, a $k$-truss subgraph of $G$ is the maximal subgraph, $G_k = (V_k,E_k,W_k)$, such that each edge $e_{ij} \in E_k$ belongs to at least ($k-2$) triangles.\label{def:ktruss}
\end{defn}

\begin{figure*}[ht!]
\begin{subfigure}{.5\columnwidth}
\centering
\includegraphics[scale=0.31]{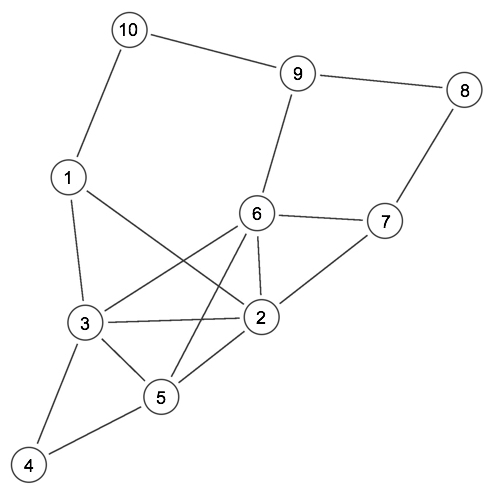}
\caption{Graph $G$}
\label{Fig:graph-G}
\end{subfigure}%
\begin{subfigure}{.5\columnwidth}
\centering
\includegraphics[scale=0.31]{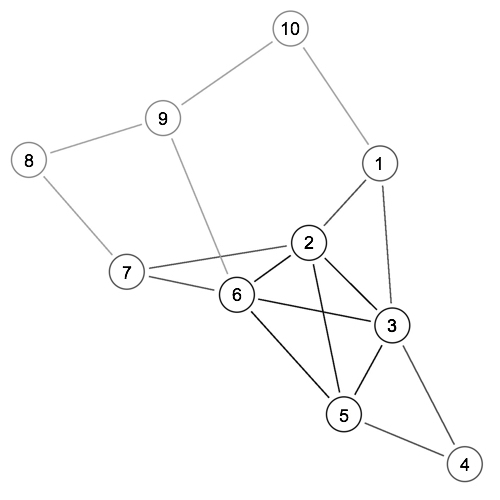}
\caption{$k$-truss subgraphs of $G$}
\label{Fig:truss-G}
\end{subfigure}
\caption{Truss-based decomposition of graph $G$ \label{Fig:truss-plotG}}
\end{figure*}

Thus, truss based decomposition results in a hierarchy of subgraphs with $G$ itself being a 2-truss graph. $G_{i+1}$ (at level ($i+1$)) is a subgraph of $G_i$ (at level ($i$)). An edge $e_{ij}$ is said to be at trussness level $l_{ij} = k$ if it lies in $k$-truss but not in $(k+1)$-truss. Higher truss level of an edge $e_{ij}$ indicates its participation in more triangles and hence, more number of common neighbors for nodes $v_i$ and $v_j$. An example graph $G$ and its $k$-trusses are shown in Figure \ref{Fig:truss-plotG}. In this example (Figure \ref{Fig:truss-G}), darker colors indicate higher truss level of the edges. Graph $G$ is decomposed into 3 subgraphs - 2-truss (graph G itself), 3-truss (graph G excluding light grey edges), and 4-truss (only dark grey edges).

Hierarchy of edges naturally translates to the hierarchy of nodes linked by the edges. Extending the concept of trussness to nodes, Kaur et al. \cite{kaur2017leveraging} define truss level $\lambda_i$ of node $v_i$ as follows.
\begin{defn}
Truss level $\lambda_i$ of node $v_i$ is defined as 
\begin{equation}
\label{def:nodeTruss}
\lambda_i =  max_{v_j \in N_i} \{l_{ij}\}
\end{equation}
where $N_i$ is the set of neighbors of node $v_i$ and $l_{ij}$ is the truss level of edge $e_{ij}$. 
\end{defn}

\begin{figure}[ht!]
\centering
\begin{minipage}{.7\linewidth}
    \includegraphics[scale=0.5]{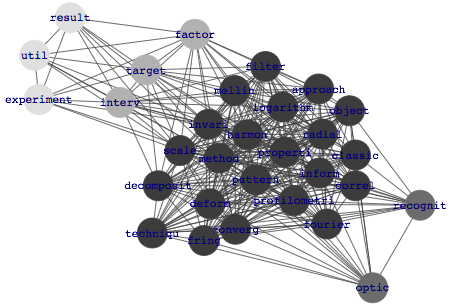}
\end{minipage}%
\begin{minipage}{.3\linewidth}
    \caption{Truss-based decomposition of CAG graph in Fig. \ref{Fig:graph-ldke}}
\label{Fig:truss-plot}
\end{minipage}
\end{figure}

Higher truss level of a node is the evidence of greater extent of its connectivity to other nodes. Figure \ref{Fig:truss-plot} shows truss-based decomposition and the corresponding node truss levels for the sample graph in Figure \ref{Fig:graph-ldke}. Different colors indicate different truss levels, with darker colors representing higher truss level of the nodes. The graph is decomposed into 4 subgraphs\footnote{All intermediate k-truss subgraphs are same. For example, in Figure \ref{Fig:truss-plot} 10-, 11-, and 12-truss subgraphs are same.} - 9-truss (the graph itself), 12-truss (the graph G excluding light grey nodes), 16-truss (graph G induced by two darker shades), and 22-truss (graph induced by darkest shade).

In the context of text graph, existence of a node at a particular truss level indicates the hierarchy level at which the word (node) $v_i$ is embedded in the text. Thus the truss levels of the nodes depict contextual hierarchy of the words in text. SC-based scoring method recognizes truss level $\lambda_i$ of a node as a factor that determines the importance of the word.

To determine the $k$-truss subgraphs of $G$, a naive algorithm iteratively removes those edges which are not part of $(k-2)$ triangles (Please see Cohen~\cite{cohen2008trusses} for details). The algorithm has a polynomial time complexity and is bounded above by $(nm^2 + n)$, where $n$ is the number of vertices and $m$ is the number of edges in $G$ \cite{cohen2008trusses}. Wang et al. proposed an algorithm for in-memory truss-based decomposition of the graph, which has time complexity $O(m^{1.5})$ and space complexity $O(n+m)$ \cite{wang2015community}. We implement this algorithm to perform truss-based decomposition of context-aware graphs in our experiments.

\subsection{Semantic Strength of a Word}
Importance of a word in the text is a function of (i) the strength of its semantic relationship with other words co-occurring in the same context, and (ii) the level of these words in contextual hierarchy. Strength of relationship between two words $v_i$ and $v_j$ is marked by the number of times the two words co-occur in same context, and is captured by weight $w_{ij}$ of edge $e_{ij}$. The semantic strength of a word (node) is defined as follows.%the sum of the strength of its relationship with neighbors weighted by their respective truss levels. Formally, %We introduce the semantic strength of a word in the text as follows.
\begin{defn}
For a node $v_i$ with neighborhood $N_i$ in graph $G$, the semantic strength of $v_i$ is defined as 
\begin{equation}
\label{def:cloutIndex}
\chi_i = \sum_{v_j \in N_i} w_{ij} \times \lambda_j
\end{equation}
where $w_{ij}$ is the weight of edge $e_{ij}$ and $\lambda_j$ is the truss-level of $v_j \in N_i$. 
\end{defn}

According to Equation \ref{def:cloutIndex}, semantic strength of word $w$ is the additive function of the co-occurrence frequency with its neighbors and their respective hierarchical levels. A word gains strength when it co-occurs frequently with other words at higher levels of hierarchy.

\subsection{Semantic Connectivity}
A document comprises multiple concepts that are semantically related. {\it KeyGraph} algorithm proposed by Ohsawa et al. finds important terms that hold the rest of the document together via inter-term connectivity between the concepts manifesting as clusters in a text graph~\cite{ohsawa1998keygraph}. We extend this idea to quantify importance of a word by counting the number of concepts in which it participates. 

In order to avoid computationally expensive task of graph clustering ($O(m^2)$ in ~\cite{ohsawa1998keygraph}), we use truss as proxy for  cluster (concept). The assumption is reasonable since clusters and trusses both highlight denser regions of the graph. Thus truss-based decomposition of the text graph yields not only the position of words in the hierarchy, but also the hierarchy of concepts. Experimental results presented in Section \ref{subsec:comp-eval-scores} validate this assumption. 
 
The extent of semantic connectivity of a word is measured by examining the number of distinct concepts that it links. If more of its neighbors belong to different concepts, its removal is likely to leave bigger semantic gap in the document. On the other hand, if all neighbors of a word belong to the same concept, removal of the word leads to little loss of meaning since the semantic relation among remaining words in the concept remains more or less intact.

Based on this premise, we approximate semantic connectivity of a word by examining the set of co-occurring words to ascertain the number of distinct concepts it links. Semantic Connectivity $SC_i$ of node $v_i$ is the count of distinct concepts (hierarchy levels) to which its neighbors belong, normalized by the highest hierarchy level in the graph. We express this measure as follows.
\begin{defn}
For a node $v_i\in G$ with neighborhood $N_i$, the semantic connectivity index of $v_i$ is defined as 
\begin{equation}
\label{def:SC}
SC_i = \frac{|\{\lambda_k: v_k \in N_i\}|}{maxtruss} 
\end{equation}
where $maxtruss$ is the highest truss (hierarchy) level of $G$.
\end{defn}

Thus, a word connected to more words from different levels of hierarchy (truss) binds together more concepts in the text, and is considered important.

\subsection{Positional weight}
Previous studies hypothesize that keywords tend to occur at the beginning of the document \cite{florescu2017position, hulth2003improved, zhang2007comparative}. PositionRank \cite{florescu2017position} is a recent development which capitalizes on this assumption to identify keywords, and is found to be an improvement over the previous methods.
Following  this premise, we take  the positional weight of each word into account while computing the word score. As prescribed by \cite{florescu2017position}, each term $t_i$ is assigned a weight based on the positional information as follows. 
\begin{equation}
    \omega_i = \sum_{j}^{n_i} \frac{1}{p_j}
\end{equation}
where $n_i$ is the frequency of term $t_i$ and $p_j$ is the $j^{th}$ position of its occurrence in the document. 

Thus, words occurring towards the beginning of the text documents are considered better candidates for keywords and are assigned higher weight than those occurring towards the end of the text document.

\subsection{Word Score}
Overall relevance of a word in the document is a function of the level at which the word is embedded ($\lambda$), semantic strength it derives from its co-occurring words ($\chi$), extent to which it is linked to the concepts present in the document ($SC$), and its positional weight ($\omega$). Assuming that these factors have a multiplicative effect on the relevance of the word in a document, word score of the candidate keyword (node) $v_i$ is defined as follows.
\begin{equation}
\label{eq:scscore}
    SCScore(v_i) = \lambda_i*\chi_i*SC_i*\omega_i
\end{equation}

Admittedly, more sophisticated functions for word scoring can be designed and explored empirically. We choose to go with  Eq. \ref{eq:scscore} because of its computational efficiency, simplicity and ease of interpretation. 

Experimental evaluation establishes effectiveness of  the proposed scoring function. For the example text in Figure \ref{Fig:text}, PositionRank is not able to extract words like ``logarithmic" and ``invariant" as the positional weights of these words pull down their ranks. On the other hand, sCAKE correctly extracts these words because it takes into account the semantic connectivity among words. This observation establishes that positional information alone is not sufficient for weighting relevance score for words. Semantic connectivity plays an important role in identifying important words in a document.

\subsection{Empirical Evaluation of SCScore function}
\label{subsec:compWordScore}
We evaluate effectiveness of the SCScore function by applying it on four types of graphs (TG, DG, GoW, and CAG) and comparing the results using respective native scoring functions. Table \ref{tab:compSCScore} reports  F1-scores for each of the graph types on the four datasets, macro-averaged at the dataset level. For ease of comparison, we repeat the result of the corresponding native scoring methods from Table \ref{tab:perf-eval-graphs}. 
\begin{table}[ht]
\scriptsize
    \caption{\label{tab:compSCScore}F1-score obtained by applying SCScore on different graph types. TG$_{TR}$: TextRank on TG graphs, TG$_{PR}$: PositionRank on TG graphs}
    \begin{minipage}{\columnwidth}
    \begin{center}
        \begin{tabular}{ c | c c | c c | c c | c c }
        \hline
        \multirow{2}{*}{\textbf{Graph}} & \multicolumn{2}{c|}{\textbf{Hulth2003}} &
      	\multicolumn{2}{c|}{\textbf{Krapivin2009}} & 
      	\multicolumn{2}{c|}{\textbf{NML500}} &
      	\multicolumn{2}{c}{\textbf{Semeval2010}}\\\cline{2-9}
      	& native & SCScore & native & SCScore & native & SCScore & native & SCScore \\
      	\hline
      	\textbf{TG$_{TR}$} & 18.37 & \textbf{51.14} & 13.72 & 38.97 & 10.73 & 23.28 & 13.65 & 34.93\\ %\hline
      	\textbf{TG$_{PR}$} & 50.41 & \textbf{51.14} & 37.07 & 38.97 & 21.94 & 23.28 & 27.50 & 34.93\\ %\hline
      	\textbf{DG} & 18.22 & 46.55 & 13.34 & 21.24 & 10.91 & 16.01 & 14.34 & 23.07\\ %\hline 
      	\textbf{GoW} & 43.41 & 43.06 & 22.70 & 30.01 & 20.20 & 20.80 & 29.34 & 29.05\\ %\hline
       \textbf{CAG} & - & 51.09 & - & \textbf{43.52} & - & \textbf{28.29} & - & \textbf{40.14}\\ %\hline
      	\hline
        \end{tabular}
    \end{center}
    \end{minipage}
\end{table} 

We observe that the performance of SCScore is significantly superior than the native word scoring methods of the four competing algorithms. We further conclude that the combination of CAG graphs and SCScore scoring method outperforms the four state-of-the-art methods .

%%%%%%%%%%%%%%%% SCAKE %%%%%%%%%%%%%%%%%%%%
%\input{mycontent/sCAKE-vs-PositionRank}
\section{sCAKE: \textit{s}emantic Connectivity Aware Keyword Extraction}
\label{sec:sCAKEvsPR}
Having illustrated the superiority of CAG graph construction method and semantic connectivity based word scoring method individually, we now integrate the two and propose a novel automatic keyword extraction algorithm named sCAKE. Three stages of the algorithm are as follows:
\begin{enumerate}
\item Candidate Filtration: Following \cite{rousseau2015main}, we identify the candidate keywords as nouns and adjectives, retained after POS tagging\footnote{\url{https://opennlp.apache.org/docs/1.8.2/manual/opennlp.html}}. This step is followed by stopword removal\footnote{\url{http://www.lextek.com/manuals/onix/stopwords2.html}} and stemming\footnote{\url{https://tartarus.org/martin/PorterStemmer/}} of the retained list. The stemmed version of the list is considered as candidates, and is passed on to the next stage along with the stemmed version of the original text.
\item Graph Construction: We create Context-Aware Text Graphs (CAG) as described in Section \ref{sec:CAGC}. This approach captures the pragmatics of written communication and connects words that are closely related to each other depending on the context of their occurrence. Unlike other methods, the proposed graph construction method is parameter-free.
\item Word Score: We compute word score for the candidates using the proposed semantic connectivity based word scoring method (SCScore) as presented in Section \ref{sec:SCbased-word-score}. This method is based on the intuition that a word derives importance from its neighbors and its own position in the text. The SCScore method exploits the semantic connectivity between words in a document based on their contextual hierarchy. This method tries to capture the semantic aspects solely on the basis of word-to-word relation without using any linguistic tools.
\end{enumerate}

The candidates are ranked according to their respective $SCScore$, and the user can extract top-$k$ candidates.

\subsection{Illustration of Keyword Extraction by Different Methods}
\label{subsec:example-scake}
We compare the set of keywords extracted by sCAKE and the four state-of-the-art methods. Table \ref{tab:comp-keywords} shows keywords extracted by the five methods against gold-standard keywords for the example text shown in Figure \ref{Fig:text}. In this example, we measure commonality of keywords between gold-standard set and the set extracted by each of the five algorithms using Jaccard Index (JI)~\cite{jaccard1901etude}. It is evident that JI for sCAKE is highest among all the methods. % The keyword `invari' matches two of the gold-standard keywords `invariance' and `invariant'. This illustrates that stemming boosts the performance of keyword extraction methods by improving recall.
K-core performs poorly on this document because it extracts words belonging to the highest core irrespective of the number of keywords to be extracted.
\begin{table}[ht!]
\scriptsize
\caption{Comparative lists of top-27 keywords extracted by sCAKE and competing methods against 27 gold-standard keywords for text in Figure \ref{Fig:text}. $\mathbf{r}$: Number of keywords that match to gold standard, $\mathbf{JI}$ = Jaccard Index. Words in \textit{italics} do not match with gold-standard. \label{tab:comp-keywords}}
\begin{minipage}{\columnwidth}
\begin{center}
\begin{threeparttable}
\begin{tabular}{ c c c c }
\hline
\textbf{Method} & \textbf{Keywords} & $\mathbf{r}$ & $\mathbf{JI}$\\
\hline
\textbf{\makecell{Gold-\\standard}} & \makecell{optical, recognition, object, scale, invariance, classical, convergent,\\ correlator, realtime, method, information, deformed, fringe,\\ patterns, fourier, transform, profilometry, technique, property, mellin,\\ radial, harmonic, decomposition, logarithmic, filter, invariant, factors}& 27 & -\\
\hline
\textbf{sCAKE} & \makecell{optic, recognit, object, scale, invari, correl, classic,\\ converg, method, inform, radial, harmon, deform, fring,\\ pattern, fourier, profilometri, techniqu, properti, \textit{approach},\\ mellin, decomposit, logarithm, filter, \textit{target}, \textit{interv}, factor} & 24\tnote{*} & 0.80 \\
\hline
\textbf{TextRank} & \makecell{scale, invariance, \textit{objects}, \textit{d}, radial, \textit{threedimensional}, harmonic, \\fourier, patterns, mellin, method, correlator, \textit{results}, filter, factors, \\convergent, classical, logarithmic, decomposition, deformed, \\fringe, profilometry, technique, \textit{approaches}, \textit{experimental},\\ recognition, invariant} & 21 & 0.64\\
\hline
\textbf{PositionRank} & \makecell{optical, recognition, \textit{objects}, scale, invariance, classical,\\\textit{threedimensional}, convergent, correlator, method, information,\\radial, harmonic, patterns, deformed, fringe, fourier, profilometry,\\ technique, property, \textit{d}, filter, mellin, decomposition, \textit{approaches},\\ logarithmic, invariant} & 23 & 0.74\\
\hline
\textbf{K-core} & classic, method, object, \textit{three-dimension} & 3 & 0.11\\
\hline
\textbf{DegExt} & \makecell{scale, \textit{d}, \textit{objects}, convergent, harmonic, invariance, radial,\\ \textit{threedimensional},  \textit{approaches}, \textit{changes}, classical, correlator,\\ decomposition, fringe, information, invariant, logarithmic, mellin,\\ method, recognition, deformed, \textit{different}, \textit{experimental},\\ factors, filter, \textit{interval}, optical} & 19 & 0.54\\
\hline
\end{tabular}
 \begin{tablenotes}
  \item[*] The keyword `invari' matches two of the gold-standard keywords `invariance' and `invariant'.
 \end{tablenotes}
\end{threeparttable}
\end{center}
\end{minipage}
\end{table}

\subsection{Comparative Evaluation of sCAKE with PositionRank}
\label{subsec:comp-eval-scores}
We empirically evaluate and compare sCAKE with PositionRank algorithm. We choose only PositionRank for comparison with sCAKE because it outperforms other three state-of-the-art methods in the earlier experiments. We extract top $k$ candidates\footnote{$k$ refers to the values as mentioned in Section \ref{sec:experimental-setup}} as keywords and compute precision, recall, and F1-score for both methods on the four datasets. The empirical results are reported in Table \ref{tab:scakeVSpr}. Bold-faced values indicate maximum F1-score for each dataset.
\begin{table}[!htb]
\begin{minipage}{.7\linewidth}
\scriptsize
\begin{center}
      	\begin{tabular}{ c | c c c | c c c }
      	%\toprule 
      	\hline
      	\multirow{2}{*}{\textbf{\makecell{Datasets}}} & \multicolumn{3}{c}{\textbf{PositionRank}} &
      	\multicolumn{3}{c}{\textbf{sCAKE}}\\
      	%\cmidrule{3-8}
      	\cline{2-7}
      	& \textbf{P} & \textbf{R} & \textbf{F1} & \textbf{P} & \textbf{R} & \textbf{F1}\\
      	%\midrule
      	\hline
      	\textbf{Hulth2003} & 45.68 & 64.45 & 50.41 & 45.41 & 66.81 & \textbf{51.09}\\ %\hline
      	\textbf{Krapivin2009} & 36.95 & 40.90 & 37.07 & 42.48 & 48.78 & \textbf{43.52}\\ %\hline 
      	\textbf{NLM500} & 19.69 & 26.60 & 21.94 & 24.88 & 34.99 & \textbf{28.29}\\ %\hline
      	\textbf{Semeval2010} & 25.31 & 31.29 & 27.50 & 35.82 & 47.37 & \textbf{40.14}\\ %\hline
      	\hline
      	\end{tabular}
      	\end{center}
      	\end{minipage}
      	%\hspace{6cm}
      	\begin{minipage}{.25\linewidth}
      	\caption{\label{tab:scakeVSpr}Performance evaluation of sCAKE vs. PositionRank}
\end{minipage}
\end{table}

We observe that the performance of sCAKE is consistently and significantly better than PositionRank on all four datasets. The improvement for longer documents is significantly higher. It is reasonable to conclude that sCAKE extracts markedly better keywords from documents of varied length compared to the competing method.

%%%%%%%%%%%%%%%% LAKE %%%%%%%%%%%%%%%%%%%%
%\input{mycontent/LAKE-method}
\section{LAKE: Language-Agnostic Keyword Extraction}
\label{sec:lake-method}
Most of the existing automatic keyword extraction algorithms use sophisticated NLP tools, which prohibits their application to texts of languages with meager NLP support. We mitigate this problem by proposing a language agnostic keyword extractor (LAKE) for eliciting keywords from a document written in language with deficient set of NLP tools. % that lack the support of sophisticate linguistic tools. 
The method profits from the strength of statistical and graph based methods, sans the burden of linguistic tools.

Like classical graph-based keyword extraction methods, LAKE is orchestrated in three stages. It is the first stage of candidate filtration that makes LAKE unique and imparts language independence. In the following subsections, we describe the candidate filtration approach for LAKE.

\subsection{Candidate keywords selection}
\label{subsec:candidate-selection}
Unlike existing graph based keyword extraction methods that accept nouns and adjectives as candidate keywords, LAKE method identifies candidate keywords by application of a statistical filter. The only input this method uses is a stopwords list curated by the user. The statistical filter is based on the computation of $\sigma$-index proposed by Ortuno et al.~\cite{ortuno2002keyword}. The idea is based on the hypothesis that the spatial distribution of a word is prime determinant of its relevance, irrespective of its frequency. The relevance is quantified by measuring the standard deviation of the distance between successive occurrences of the word in the text \cite{ortuno2002keyword}. The use of this filter substantially reduces the search space, and imparts language independence to this stage. 

We consider Zhou et al. \cite{zhou2003metric} for implementation of $\sigma$-index, which considers boundary values for computation. Consider a word $w$ that occurs $n$ times in a document of length $N$. Let $p_i$ denote the position of $i^{th}$ occurrence of $w$, with boundary values $p_0$ and $p_{N+1}$ set to $0$ and $N+1$, respectively. Then ($p_{i+1}-p_i$) denotes the distance between two consecutive occurrences of $w$. The average distance between occurrences of $w$ is given by $\mu(w)$,
$$
\mu(w) = \frac{(p_1 - p_0) + (p_2 - p_1) + ... + (p_{N+1} - p_n)}{n+1} = \frac{N+1}{n+1},
$$
and the standard deviation is given by
$$s(w) = \sqrt{\frac{1}{n-1} \sum_{i = 0}^{n}((p_{i+1}-p_i)-\mu(w))^2}$$

The $\sigma$-index $\sigma(w)$ of $w$, is defined as
\begin{equation}
\sigma(w) = \frac{s(w)}{\mu(w)} \text{,}
\end{equation}
 
Table \ref{tab:filter_performance} shows the comparative performance of POS tag-based filter and statistical filter. The values in each cell present the overlap between the gold-standard keywords and the set of candidates obtained by each of these filters. It is evident from the table that the best candidate list is obtained after performing stemming on the list of nouns and adjectives. $\sigma$-index produces noisy candidates list, which is the price paid for language independence.
\begin{table}[ht!]
\scriptsize
%\begin{center}
\begin{minipage}{3cm}
    \begin{center}
    \begin{tabular}{ c c c c }% m{1.3cm} | m{2cm} | }
        %\toprule
        \hline
        \textbf{Dataset} & $\sigma$ & \textbf{PoST}& \textbf{PoST+stem}\\ %& \textbf{Maximum F1-score reported}\\
        %\midrule
        \hline
        \textbf{Hulth2003} & - & 85.38  & 87.33\\% & ~\cite{hulth2003improved,mihalcea2004textrank,rousseau2015main} & 51.92 \\
        %\hline
        \textbf{NLM500} & 53.12 %& 53.09 
        & 69.54 & 76.90\\% & a & 1 \\
        %\hline
        \textbf{Krapivin2009} & 88.13 & 92.94 %& 88.09 
        & 95.45\\% & ~\cite{rousseau2015main} & 50.77 \\
        %\hline
        \textbf{SemEval2010} & 80.55 & 89.93 %& 80.54 
        & 94.45\\% & a & 1 \\
        %\bottomrule
        \hline
        \end{tabular}
        %\end{center}
        \end{center}
\end{minipage}
\hspace{3cm}
\begin{minipage}{5.5cm}
    \caption{Percentage  overlap of gold-standard keywords and candidate lists obtained by: $\sigma$: $\sigma$-index, PoST:  retaining nouns and adjectives using POS tagging, stem: with stemming enabled.\label{tab:filter_performance}}
\end{minipage}
\end{table}

\begin{figure}[ht!]
\centering
\begin{subfigure}{\textwidth}
\centering
%\begin{minipage}{4cm}
    \includegraphics[scale=0.45]{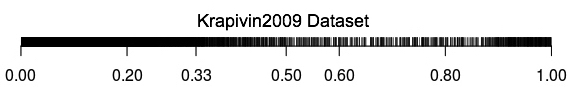}
    %\caption{Krapivin2009 dataset}
    \label{Fig:sigkrapi}
%\end{minipage}
\end{subfigure}
\begin{subfigure}{\textwidth}
\centering
%\begin{minipage}{4cm}
    \includegraphics[scale=0.45]{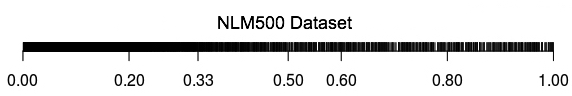}
    %\caption{NLM500 dataset}
    \label{Fig:signlm}
%\end{minipage}
\end{subfigure}
\begin{subfigure}{\textwidth}
\centering
%\begin{minipage}{4cm}
    \includegraphics[scale=0.45]{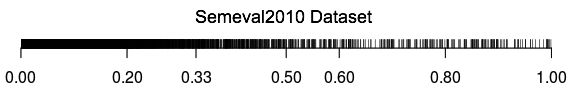}
    %\caption{Semeval2010 dataset}
    \label{Fig:sigsem}
%\end{minipage}
\end{subfigure}
\caption{Distribution of normalized $\sigma$-index ranks of gold-standard keywords contained in the candidate lists}
\label{Fig:rugplotSigma}
\end{figure}

In order to find the threshold $\sigma$-index for retaining candidate keywords, we inspected the ranks of gold-standard keywords by $\sigma$-index scores. Rugplots in Figure \ref{Fig:rugplotSigma} show higher density in the region corresponding to higher ranks (lower values correspond to higher ranks). We found that on an average, more than 92$\%$ gold-standard keywords out of those occurring explicitly in the text\footnote{Some gold-standard keywords do not appear in the text.} occur in the top 33$\%$ words ranked using $\sigma$-index. Based on this observation, we decide to retain top-33\% candidates ranked based on $\sigma$-index. Since $\sigma$-index is not suitable for small length documents, we do not apply this filter for Hulth2003 dataset in the experiments reported in Section \ref{subsec:lake-vs-scake}.

It is evident from Figure \ref{Fig:rugplotSigma} that gold-standard keywords are usually ranked higher based on their corresponding $\sigma$-index. For NLM500 dataset, the $\sigma$-index based ranks of gold-standard keywords tend to gather towards top-33$\%$ with anomalies lying towards lower ranks. This affects the performance for NLM500 dataset, which is reflected in the empirical results.

\subsection{Experimental Evaluation: LAKE vs. sCAKE}
\label{subsec:lake-vs-scake}
We compare the performance of LAKE with sCAKE to assess the amount of performance degradation due to non-adoption of NLP tools in LAKE method. 
\begin{table}[!htb]
	\scriptsize
     \caption{\label{tab:lakeVSscake}Performance evaluation of LAKE vs. sCAKE}
\begin{center}
      	\begin{tabular}{ c | c c c | c c c | c }
      	%\toprule 
      	\hline
      	\multirow{2}{*}{\textbf{\makecell{Datasets}}} & \multicolumn{3}{c}{\textbf{sCAKE}} &
      	\multicolumn{3}{c|}{\textbf{LAKE}} & \multirow{2}{*}{\textbf{\makecell{Performance\\Loss in F1 ($\%$)}}}\\
      	%\cmidrule{3-8}
      	\cline{2-7}
      	& \textbf{P} & \textbf{R} & \textbf{F1} & \textbf{P} & \textbf{R} & \textbf{F1}\\
      	%\midrule
      	\hline
      	\textbf{Hulth2003} & 45.41 & 66.81 & \textbf{51.09} & 41.67 & 59.31 & 46.14 & 9.69\\ %\hline
      	\textbf{Krapivin2009} & 42.48 & 48.78 & \textbf{43.52} & 37.60 & 41.56 & 37.69 & 13.4\\ %\hline 
      	\textbf{NLM500} & 24.88 & 34.99 & \textbf{28.29} & 19.60 & 26.55 & 21.87 & 22.69\\ %\hline
      	\textbf{Semeval2010} & 35.82 & 47.37 & \textbf{40.14} & 29.48 & 36.48 & 32.08 & 20.08\\ %\hline
      	\hline
      	\end{tabular}
      	\end{center}
\end{table}

It is evident from the above table (Table \ref{tab:lakeVSscake}) that there is a considerable performance gap between NLP-enabled and language-agnostic variation. For English-like languages that enjoy the support of sophisticated NLP tools, sCAKE is a better choice as it outperforms the other state-of-the-art keyword extraction methods. However, for languages that lack the support of sophisticated NLP tools, there is no alternative approach provided by the existing methods to enable language-independence feature. Thus, LAKE seems to be a fair solution which can be applied on languages without linguistic support, albeit with an associated cost of performance degradation.

\subsection{Comparative Evaluation of Competing Methods}
\label{subsec:comp-graph}
Figures \ref{Fig:comp-graph-all}(a-d) show comparative line graphs for sCAKE, LAKE, PositionRank, and K-core methods per dataset. It is evident that sCAKE (red opaque diamond line) outperforms all other methods. Performance of LAKE is at par with PositionRank, outperforming K-core in all four datasets. F1-score for K-core does not improve with increasing keywords because K-core always extracts words belonging to the top-most core as keywords. {As stated earlier, F1-score for all methods drop for very low and very high number of keywords. This is because for less number of keywords, precision is usually high but recall is low. On the other hand, for very large number of keywords, recall is high but precision is low. This ultimately affects the F1-score, bringing it down to a lower value.

\begin{figure}[ht]
\centering
\begin{subfigure}{.5\textwidth}
  \centering
\includegraphics[scale=0.25]{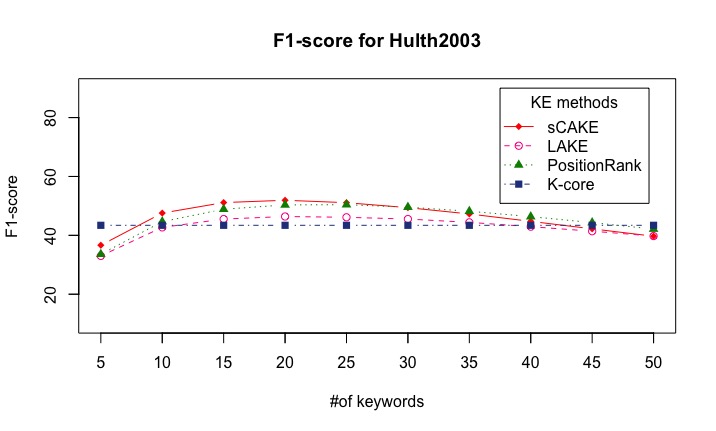}
\caption{Hulth2003 dataset}
\label{Fig:comp-graph-hulth}
%\bigskip
\end{subfigure}%
%\hspace{0.5cm}
\begin{subfigure}{.5\textwidth}
  \centering
\includegraphics[scale=0.25]{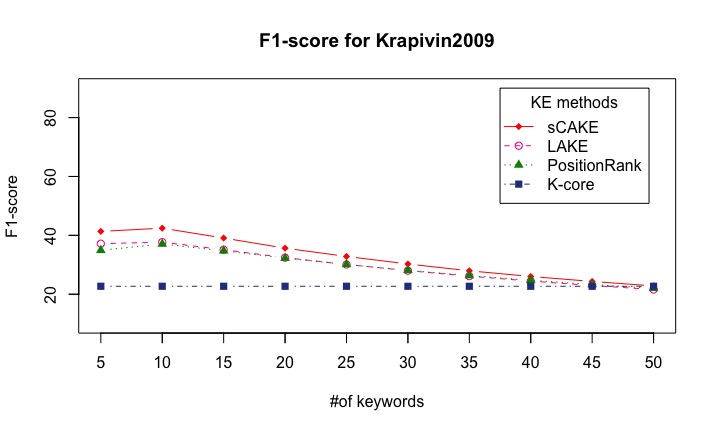}
\caption{Krapivin2009 dataset}
\label{Fig:comp-graph-krapi}
%\bigskip
\end{subfigure}
\begin{subfigure}{.5\textwidth}
  \centering
\includegraphics[scale=0.25]{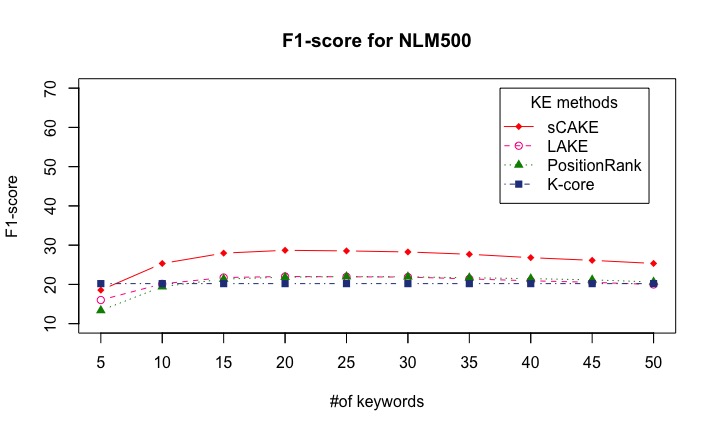}
\caption{NLM500 dataset}
\label{Fig:comp-graph-nlm}
%\bigskip
\end{subfigure}%
%\hspace{0.5cm}
\begin{subfigure}{.5\textwidth}
  \centering
\includegraphics[scale=0.25]{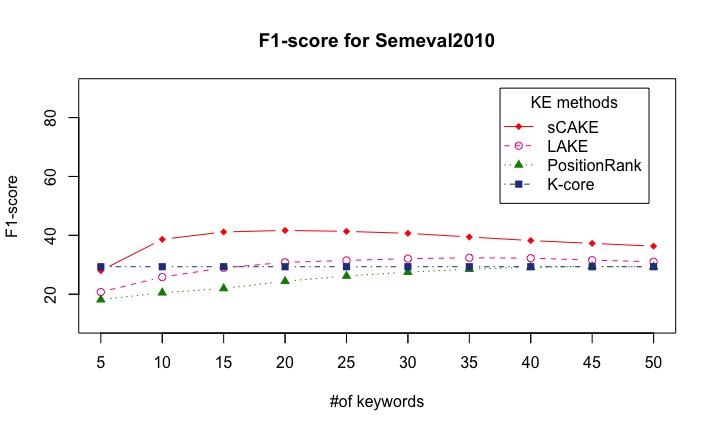}
\caption{Semeval2010 dataset}
\label{Fig:comp-graph-semeval}
%\bigskip
\end{subfigure}
\caption{Lineplot of F1-score for sCAKE, LAKE, PositionRank, and $k$-core on each dataset.}
\label{Fig:comp-graph-all}
\end{figure}

\subsection{Experimentation on Indian Languages}
\label{subsec:LAKE-on-assamese}
India is a country with 23 official languages, including English. According to Census of India of 2001, India has 122 major languages and 1599 other languages. With such a wide variety of written and spoken languages, there is a huge collection of literature available. However, due to scarcity of sophisticated NLP tool, automatic analysis of such documents is challenging.

We evaluated LAKE method for automatic keyword extraction from an Wikipedia article on `Animation' written in Assamese. We removed English characters from this document as an additional pre-processing step. The stopwords list used for this exercise is downloaded from TDIL website\footnote{\url{https://www.tdil-dc.in/index.php?option=com_download&task=showresourceDetails&toolid=1634&lang=en}}. Top-10 extracted keywords, with their respective translations to English, are shown in Table \ref{tab:as-keywords}.

\begin{table}[!htb]
	\scriptsize
     \caption{\label{tab:as-keywords}Sample keywords extracted from Assamese text}
\begin{center}
      	\begin{tabular}{ c }
      	%\toprule 
      %	\begin{figure}[ht!]
%\centering
    \includegraphics[scale=0.5]{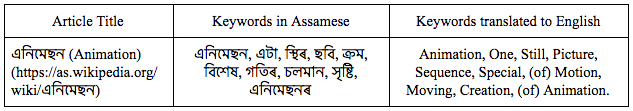}
    %\caption{Sample keywords extracted from Assamese text}
    %\label{Fig:as-keywords}
%\end{figure}
      	\end{tabular}
      	\end{center}
\end{table}

The document along with the set of programs and stopwords list are available at GitHub\footnote{\url{https://github.com/SDuari/LAKE-on-Assamese-text}}. Due to non-availability of gold-standard keywords set, we could not evaluate the performance of LAKE on Assamese text. We leave it to the readers to judge the performance based on the extracted keywords.

%%%%%%%%%%%%%%%% CONCLUSION %%%%%%%%%%%%%%%%%%%%
%\input{mycontent/Conclusion}
\section{Conclusion}
\label{sec:conclusion}
We present a commentary on graph-based keyword extraction methods, and propose two new parameter-free methods sCAKE and LAKE. The two methods are based on novel sentence-based graph construction approach (CAG) that is mindful of the carriage of pragmatics from each sentence to its following one. The novel word scoring approach (SCScore) computes the relevance of words by taking into account its contextual hierarchy, semantic connectivity, and positional weight in the text.

We first evaluate the proposed graph construction and word scoring methods individually, and subsequently integrate as sCAKE algorithm. Four state-of-the-art keyword extraction methods - TextRank, DegExt, $k$-core Retention, and PositionRank were compared using four benchmark datasets. Experimental results reveal that the native word scoring methods perform better on CAG graphs compared to the corresponding graphs. We also observe that the proposed word scoring method performs consistently better than other scoring methods irrespective of the graph construction approach. Further, we show that the proposed keyword extraction method sCAKE outperforms PositionRank in terms of F1-score.

A language-agnostic variant of sCAKE (called LAKE) is proposed which  employs statistical filter to identify candidate keywords. As expected, LAKE suffers performance degradation compared to sCAKE on the studied datasets, all of which consists of English texts. We conclude that for languages with sophisticated NLP support, it is better to exploit the linguistic features. However, LAKE method  can be applied on languages that are not supported with sophisticated NLP tools, albeit with an associated cost of performance degradation.

Top-10 keywords extracted (after stemming) by sCAKE method from this manuscript\footnote{Excluding conclusion, references, and other non-text entities like tables and figures with captions} are - ``keyword", ``scake", ``extract", ``semant", ``connect", ``method", ``text", ``awar", ``graph", and ``word". All the words in the title are included in the top-10 keywords list, which is desirable.

In future, we intend to apply LAKE on documents written in Indian Languages to see how well it performs on multiple languages and domains. We also intend to make LAKE a benchmark, on the basis of which future keyword extraction algorithms for Indian languages could be tested upon.

%%%%%%%%%%%%%%%% ACKNOWLEDGE %%%%%%%%%%%%%%%%%%%%
%\input{mycontent/Acknowledgement}
\section*{Acknowledgement}
\label{sec:ack}
The authors acknowledge the financial support (Grant number RC/2015/9677) awarded by University of Delhi, India for this research.

\section*{References}

%\newpage

\bibliography{sCAKE-bib}

\end{document}